\documentclass[aps,prper,amsmath,amssymb,nofootinbib,12pt]{revtex4-2}

\usepackage{amsmath,amssymb,amsfonts}
\usepackage{graphicx}
\usepackage{physics}
\usepackage{bm}
\usepackage{natbib}
\usepackage{tikz}
\usetikzlibrary{calc, arrows.meta, positioning}
\usepackage{xcolor}
\usepackage{setspace}
\usepackage{hyperref}

\begin{document}
	
	\title{Gravitational Lensing as an Optical Framework for Modified Gravity Theories}
	
	\author{Romy Hanang Setya Budhi}
	\email{romyhanang@ugm.ac.id}
	\affiliation{Faculty of Mathematics and Natural Sciences, Universitas Gadjah Mada, Yogyakarta, Indonesia}
	
	\begin{abstract}
		We present a framework that reformulates gravitational lensing as an optical phenomenon governed by an effective refractive index, enabling exploration of modified gravity theories using undergraduate-level mathematics and optics. After deriving the general deflection angle for arbitrary spherically symmetric fields, we establish the observational baseline using standard general relativity, including the lens equation and Einstein ring properties. Assuming the optical relation holds for modified effective potentials, we apply the formalism to deep-MOND, Yukawa-type, and power-law ($f(R)$) models, providing closed-form analytical expressions for the deflection angle and Einstein radius. Numerical ray-tracing simulations validate these analytical results. This framework serves as a conceptual bridge to contemporary research, offering students computational experience and critical awareness of gravitational lensing foundations.
		
		\vspace{0.5cm}
		\noindent
		\textbf{Keywords}:gravitational lensing, modified gravity, optical-mechanical analogy, Einstein ring. 
	\end{abstract}
	
	\maketitle
	
	\section{Introduction}
	
	The teaching of gravitation at the undergraduate level has traditionally centered on Newtonian mechanics, leaving students unprepared to engage with the revolutionary developments in gravitational physics over the past century. General relativity requires mathematical tools such as tensor calculus and differential geometry that typically lie beyond undergraduate curricula \cite{einstein1916,misner1973}. Consequently, students often graduate with an understanding of gravity that effectively ended in the nineteenth century, despite the fact that gravitational physics remains one of the most active frontiers of contemporary research.
	
	The discovery of the accelerated expansion of the universe \cite{riess1998,perlmutter1999} and the anomalous rotation curves of galaxies \cite{rubin1980} have spawned intense theoretical activity aimed at understanding whether these observations signal the existence of dark matter and dark energy, or whether Einstein's theory of gravity requires modification on cosmological scales. Modified gravity theories now constitute a major research paradigm, with thousands of papers published annually exploring alternatives to general relativity \cite{capozziello2011,clifton2012}. Students who lack exposure to these developments are ill equipped to understand the motivations behind current research or to contribute to this vibrant field. The challenge lies in finding an accessible entry point that bridges the gap between standard undergraduate physics and these contemporary research questions without requiring graduate level mathematical preparation.
	
	Gravitational lensing, the deflection of light by gravitational fields, offers a uniquely suitable entry point for bridging this divide. The phenomenon has a distinguished history dating back to Einstein's 1916 calculation of light deflection by the sun \cite{einstein1916}, which was famously confirmed by Eddington's 1919 expedition \cite{eddington1919,dyson1920}. Today, gravitational lensing has evolved into a sophisticated observational tool used to map dark matter distributions \cite{clowe2006}, measure the Hubble constant \cite{suyu2017}, and search for extrasolar planets \cite{mao1991}. The key insight that enables a tractable treatment of gravitational lensing is the recognition that, within the weak field limit of general relativity, light propagation in a gravitational field can be equivalently described as propagation through an optical medium with a spatially varying refractive index \cite{einstein1916,binney2008}. This equivalence, sometimes called the optical mechanical analogy, transforms the problem from the language of curved spacetime into the language of geometrical optics, a subject with which undergraduate students typically have prior exposure. The optical formulation rests on the physical picture where as light passes through a region containing a gravitational potential $\Phi(r)$, it effectively experiences a medium with refractive index $n(r) = 1 - 2\Phi(r)/c^2$. This expression emerges naturally from the weak field approximation to general relativity. The resulting mathematical framework requires only multivariable calculus and elementary differential equations, tools that physics majors typically acquire in their first two years of study.
	
	Several authors have recognized the instructional value of this optical analogy. The classic textbook by Schneider, Ehlers, and Falco \cite{schneider1992} provides a rigorous foundation at a graduate level, while more accessible treatments have appeared in educational journals \cite{nandi1995,narayan1996}. Recent studies have explored modified gravity signatures through weak lensing observations \cite{schmidt2008,terukina2011}, and parallel educational efforts have introduced modified gravity through other classical phenomena such as perihelion precession \cite{budhi2025ejp}. However, existing treatments typically focus on standard relativity or isolated phenomena without extending the lensing framework to a broader range of modified gravity theories.
	
	The present work makes two novel contributions. First, we provide a self-contained derivation of the optical formalism accessible to advanced undergraduates, starting from the fundamental inability of Newtonian gravity to predict light deflection and culminating in the general relativistic lens equation and Einstein radius. Second, we extend this framework to several modified gravity models, including MOND, Yukawa type corrections, and power law $f(R)$ modifications, showing how each theory produces a distinct optical signature. This extension transforms a standard derivation into an exploration of frontier research, bridging the gap between undergraduate education and contemporary physics.
	
	The remainder of this paper is organized as follows. Section II establishes the theoretical foundations, tracing the path from Newtonian optics to the relativistic lens equation. Section III applies this formalism to modified gravity models, deriving deflection angles and Einstein radii for each. Section IV describes the computational implementation of ray tracing simulations. Section V presents results and discussion, including scaling relations and observational constraints. Section VI concludes with implications and suggestions for further exploration.
	
	\section{Theoretical Foundations: From Newtonian Optics to the Relativistic Lens Equation}
	
	The theoretical foundation of gravitational lensing rests on a subtle but critical point that pure Newtonian gravity fundamentally cannot predict the deflection of light. This section addresses this issue directly, then develops the full optical and geometrical apparatus required to describe lensing in the weak field limit of general relativity, and finally establishes the empirical baseline against which modified gravity theories must be compared. The presentation proceeds from first principles through the derivation of the ray equation in a refractive medium, then to the general expression for the deflection angle in a spherically symmetric potential, and concludes with the thin lens geometry that connects microscopic deflections to macroscopic observables such as the Einstein ring.
	
	\subsection{Gravitational Lensing from Newtonian to General Relativistic Perspectives}
	
	In Newton's law of universal gravitation, the force on a test particle of mass $m$ due to a point source of mass $M$ is $\mathbf{F} = -GMm\hat{\mathbf{r}}/r^2$, implying that a photon, having zero rest mass ($m_\gamma = 0$), experiences no gravitational force in this framework. Consequently, light travels in straight lines through gravitational fields with no deflection whatsoever. This represents a fundamental theoretical impasse that cannot be resolved within the Newtonian framework, regardless of any auxiliary assumptions one might introduce.
	
	Historically, several physicists attempted to calculate light deflection using Newtonian concepts combined with the corpuscular theory of light, treating photons as massive particles moving at speed $c$. Soldner in 1801 \cite{soldner1804} obtained a deflection angle $\alpha_\text{Soldner} = 2GM/(bc^2)$, which is exactly half the correct value. Einstein himself initially obtained this same result in 1911 using the equivalence principle \cite{einstein1911}, but this calculation was incomplete because it accounted only for the effect of gravitational time dilation (the temporal part of the metric) and omitted the contribution from spatial curvature.
	
	The correct prediction of gravitational lensing requires the full framework of general relativity. In the weak-field limit, the spacetime metric around a spherically symmetric mass takes the form
	\begin{equation}
		ds^2 = -\left(1 + \frac{2\Phi}{c^2}\right)c^2 dt^2 + \left(1 - \frac{2\Phi}{c^2}\right)(dx^2 + dy^2 + dz^2),
	\end{equation}
	where $\Phi = -GM/r$ is the Newtonian gravitational potential. Light follows null geodesics ($ds^2 = 0$), and the coordinate velocity of light is modified by both temporal and spatial metric components:
	\begin{equation}
		v = \frac{dr}{dt} = c\sqrt{\frac{1 + 2\Phi/c^2}{1 - 2\Phi/c^2}} \approx c\left(1 + \frac{2\Phi}{c^2}\right),
	\end{equation}
	where the approximation holds to first order in $\Phi/c^2$. This dual modification gives rise to an effective refractive index
	\begin{equation}
		n(r) = \frac{c}{v} = 1 - \frac{2\Phi(r)}{c^2}.
		\label{eq:refractive_index}
	\end{equation}
	The factor of 2 in this expression is crucial because it leads to the full general relativistic prediction $\alpha_\text{GR} = 4GM/(bc^2)$, which is twice the value obtained by Soldner and early Einstein. This factor of 2 was famously confirmed by Eddington's 1919 solar eclipse observations \cite{dyson1920}, establishing general relativity as the correct theory of gravitation.
	
	\subsection{Fermat's Principle and the Ray Equation in an Inhomogeneous Medium}
	
	The foundation of geometrical optics is Fermat's principle, which states that light travels between two points along a path that renders the optical path length stationary \cite{born1999}. Mathematically, for propagation through a medium with position-dependent refractive index $n(\mathbf{r})$, this principle is expressed as
	\begin{equation}
		\delta \int_A^B n(\mathbf{r}) \, ds = 0,
	\end{equation}
	where $ds = |d\mathbf{r}|$ is the element of Euclidean arc length along the path, and the integral extends from the initial point $A$ to the final point $B$. This variational principle constitutes the optical analogue of Maupertuis's principle in classical mechanics, with the refractive index playing a role analogous to momentum.
	
	To derive the differential equation governing light rays, we parametrize the path by its arc length $s$ itself, which satisfies $|d\mathbf{r}/ds| = 1$ at every point. The optical path length becomes $S = \int_A^B n(\mathbf{r}(s)) \, ds$. Under a small variation $\mathbf{r}(s) \to \mathbf{r}(s) + \delta\mathbf{r}(s)$ with fixed endpoints $\delta\mathbf{r}(A) = \delta\mathbf{r}(B) = 0$, the change in $S$ to first order is
	\begin{equation}
		\delta S = \int_A^B \left[\nabla n \cdot \delta\mathbf{r} \, ds + n \, \delta(ds) \right].
	\end{equation}
	From the definition $ds = \sqrt{d\mathbf{r} \cdot d\mathbf{r}}$, we find $\delta(ds) = \hat{\mathbf{t}} \cdot d(\delta\mathbf{r})/ds \, ds$, where $\hat{\mathbf{t}} = d\mathbf{r}/ds$ is the unit tangent vector. Substituting and integrating by parts yields
	\begin{equation}
		\delta S = \int_A^B \left[\nabla n - \frac{d}{ds}\left(n \hat{\mathbf{t}}\right) \right] \cdot \delta\mathbf{r} \, ds + \left[ n \hat{\mathbf{t}} \cdot \delta\mathbf{r} \right]_A^B.
	\end{equation}
	The boundary term vanishes due to the fixed endpoints, and since $\delta\mathbf{r}(s)$ is arbitrary for interior points, the stationarity condition $\delta S = 0$ yields the ray equation
	\begin{equation}
		\frac{d}{ds}\left( n \frac{d\mathbf{r}}{ds} \right) = \nabla n.
		\label{eq:ray_equation_exact}
	\end{equation}
	This second-order differential equation is the fundamental equation of geometrical optics \cite{born1999}.
	
	For gravitational lensing, the refractive index deviates only slightly from unity because $|\Phi|/c^2 \ll 1$. Writing $n(\mathbf{r}) = 1 + \delta n(\mathbf{r})$ with $|\delta n| \ll 1$, we expand Eq.~(\ref{eq:ray_equation_exact}) to first order. Since $|dn/ds| \sim |\nabla n| \sim |\delta n|/L$ where $L$ is a characteristic length scale, and $n \approx 1$, the ray equation simplifies to
	\begin{equation}
		\frac{d^2\mathbf{r}}{ds^2} = \nabla n,
		\label{eq:ray_equation_weak}
	\end{equation}
	where we have dropped terms of order $\delta n \times (d^2\mathbf{r}/ds^2)$ as second-order small. This equation has the intuitive interpretation that the light ray experiences an acceleration proportional to the gradient of the refractive index, analogous to Newton's second law with $\nabla n$ playing the role of a force per unit mass.
	
	\subsection{General Expression for the Deflection Angle in Spherical Symmetry}
	
	We now specialize to a spherically symmetric gravitational field, where the potential $\Phi(r)$ depends only on the radial distance from the center of the gravitating body \cite{perlick2004}. Consider a light ray approaching from infinity with impact parameter $b$, defined as the perpendicular distance from the center to the initial direction of the ray. In the weak-deflection approximation, the actual trajectory deviates only slightly from a straight line, a condition satisfied for all astrophysically relevant lensing scenarios except those involving black holes or neutron stars. We therefore replace the true curved trajectory with the unperturbed straight line, parametrized as
	\begin{equation}
		r(s) = \sqrt{b^2 + s^2},
	\end{equation}
	where $s$ is the coordinate along the unperturbed path measured from the point of closest approach, ranging from $s = -\infty$ (approaching from far away) to $s = +\infty$ (receding to infinity).
	
	The ray equation in the weak-field limit, Eq.~(\ref{eq:ray_equation_weak}), states that the acceleration of the light ray equals the gradient of the refractive index. Decomposing this into components parallel and perpendicular to the unperturbed path, only the perpendicular component contributes to the net deflection. For a spherically symmetric potential, the perpendicular component of $\nabla n$ points radially inward and has magnitude $(dn/dr)(b/r)$. The transverse acceleration is therefore
	\begin{equation}
		a_\perp = \frac{d^2 r_\perp}{ds^2} = \frac{dn}{dr}\frac{b}{r}.
	\end{equation}
	The total transverse velocity acquired by the light ray equals the deflection angle in the small-angle approximation (since the initial transverse velocity is zero). Integrating along the entire path gives
	\begin{equation}
		\alpha = \int_{-\infty}^{+\infty} \frac{d^2 r_\perp}{ds^2} ds = \int_{-\infty}^{+\infty} \frac{dn}{dr}\frac{b}{r} ds.
		\label{eq:deflection_general}
	\end{equation}
	Substituting the relation $n = 1 - 2\Phi/c^2$ from Eq.~(\ref{eq:refractive_index}) yields
	\begin{equation}
		\alpha = -\frac{2}{c^2}\int_{-\infty}^{+\infty} \frac{d\Phi}{dr}\frac{b}{r} ds.
		\label{eq:deflection_potential_gradient}
	\end{equation}
	This formula is completely general for spherically symmetric potentials within the weak-deflection approximation. For many applications, it is convenient to express the deflection angle directly in terms of the potential rather than its gradient. Integrating Eq.~(\ref{eq:deflection_potential_gradient}) by parts (or equivalently, using Gauss's theorem in two dimensions) gives the equivalent expression
	\begin{equation}
		\alpha = \frac{2}{c^2}\int_{-\infty}^{+\infty} \Phi(r) \frac{b}{r^2} ds.
		\label{eq:deflection_potential_direct}
	\end{equation}
	This alternative form is particularly useful when the potential has a simple closed form but its gradient introduces complications, as occurs for logarithmic potentials.
	
	For the specific case of the Newtonian potential $\Phi = -GM/r$, we can evaluate Eq.~(\ref{eq:deflection_potential_gradient}) directly. Using $dn/dr = 2GM/(c^2 r^2)$ and the substitution $s = b\tan\theta$, we obtain
	\begin{equation}
		\alpha_\text{GR} = \frac{4GM}{bc^2}.
		\label{eq:alpha_gr}
	\end{equation}
	This is the celebrated result first obtained by Einstein in 1916. For light grazing the sun ($M = M_\odot$, $b = R_\odot$), this formula predicts a deflection of approximately 1.75 arcseconds. This result serves as the baseline against which modified gravity models will be compared.
	
	\subsection{Geometrical Formalism: The Thin-Lens Equation and Einstein Radius}
	
	To connect the microscopic deflection angle $\alpha$ with observable quantities on the celestial sphere, we employ the thin-lens approximation \cite{schneider1992,weinberg1972}. This approximation assumes that the physical extent of the gravitational lens along the line of sight is negligible compared to the distances separating the observer, the lens, and the background source. Consequently, the deflection is treated as occurring instantaneously at the lens plane.
	
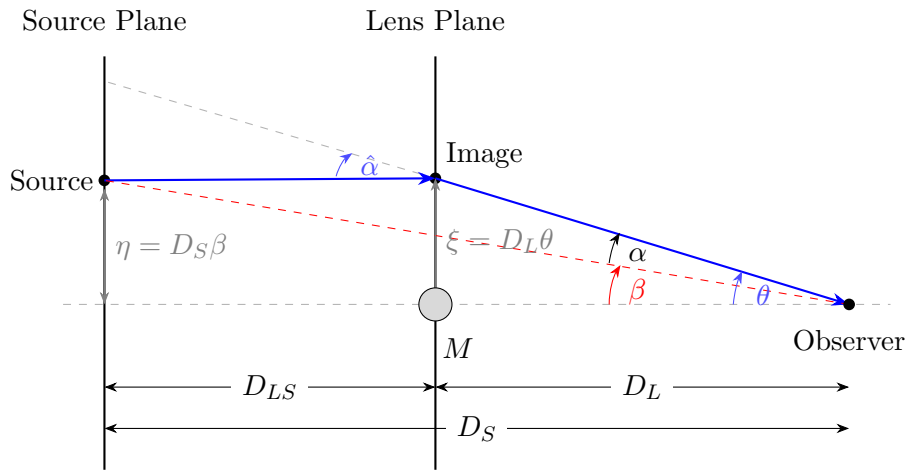
\begin{figure}[htbp]
		\centering
		\begin{tikzpicture}[
			scale=1.1,
			>=Stealth,
			every node/.style={font=\small},
			plane/.style={thick, black},
			optical axis/.style={dashed, gray!60},
			light ray/.style={blue, thick, ->},
			undeflected/.style={red, dashed, thin},
			extension/.style={gray, dashed, very thin},
			angle arc/.style={->, >=Stealth, blue!70}
			]
			\coordinate (Observer) at (0,0);
			\coordinate (LensPlane) at (-5,0);
			\coordinate (SourcePlane) at (-9,0);
			\coordinate (Source) at (-9,1.5);
			\coordinate (Image) at (-5,1.525);
			\coordinate (Image_2) at (-9,2.7);
			
			\draw[<->, black] (0,-1.0) -- (-5,-1.0) node[midway, fill=white] {$D_L$};
			\draw[<->, black] (-5,-1.0) -- (-9,-1.0) node[midway, fill=white] {$D_{LS}$};
			\draw[<->, black] (0,-1.5) -- (-9,-1.5) node[midway, fill=white] {$D_S$};
			
			\draw[optical axis] (Image) -- (Image_2);

			\draw[optical axis] (-9.5,0) -- (0.5,0);
			\draw[plane] (-5,-2) -- (-5,3) node[above, yshift=2mm] {Lens Plane};
			\draw[plane] (-9,-2) -- (-9,3) node[above, yshift=2mm] {Source Plane};
			
			\fill[black] (Source) circle (2pt) node[left] {Source};
			\fill[black] (Image) circle (2pt) node[above right] {Image};
			
			\draw[light ray] (Source) -- (Image);
			\draw[light ray] (Image) -- (Observer);
			\draw[undeflected] (Observer) -- (Source);
			
			\draw[angle arc] (-1.4,0) arc (175:155:1.2) node[midway, left, xshift=6mm, yshift=-1mm] {$\theta$};
			\draw[angle arc, red] (-2.9,0) arc (180:148:0.9) node[midway, left, xshift=6mm, yshift=-1mm] {$\beta$};
			\draw[angle arc, black] (-2.9,0.5) arc (173:148:0.9) node[midway, left, xshift=6mm, yshift=-1mm] {$\alpha$};
			
			\draw[angle arc] (-6.2,1.55) arc (165:125:0.5) node[midway, above right, xshift=1mm, yshift=-3mm] {$\hat{\alpha}$};
			
			\draw[<->, gray] (-5,0) -- (-5,1.525) node[midway, right] {$\xi = D_L\theta$};
			\draw[<->, gray] (-9,0) -- (-9,1.4) node[midway, right] {$\eta = D_S\beta$};
			
			\fill[black] (Observer) circle (2pt) node[below, yshift=-2mm] {Observer};
			\draw[fill=gray!30] (-5,0) circle (0.2) node[below, xshift=3mm,yshift=-3mm] {$M$};
			
		\end{tikzpicture}
		\caption{Geometry of the thin-lens approximation in gravitational lensing. The observer detects the image at angular position $\theta$, while the true source position is at $\beta$. The light ray is deflected by the reduced angle $\hat{\alpha} = (D_{LS}/D_S)\alpha$ at the lens plane. Angular diameter distances $D_L$, $D_S$, and $D_{LS}$ relate the physical transverse distances $\xi = D_L\theta$ and $\eta = D_S\beta$ via the lens equation $\beta = \theta - \hat{\alpha}(\theta)$.}
		\label{fig:lens_geometry}
	\end{figure}

	Consider the geometry shown in Fig.~\ref{fig:lens_geometry}. We define the angular diameter distances $D_L$, $D_S$, and $D_{LS}$ as the separations from observer to lens, observer to source, and lens to source, respectively. In a flat universe these distances satisfy $D_S = D_L + D_{LS}$ for small redshifts, but more generally they are defined via the angular diameter distance formula \cite{weinberg1972}. Let $\beta$ be the true angular position of the source in the absence of the lens, and $\theta$ the observed angular position of the image. The physical transverse distances at the lens and source planes are $\xi = D_L \theta$ and $\eta = D_S \beta$, respectively.
	
	From similar triangles in Fig.~\ref{fig:lens_geometry}, the physical distance from the optical axis at the source plane equals the distance at the lens plane minus the contribution from the deflection, yielding the relation $\eta = \xi - D_{LS} \tan\hat{\alpha}$. In the small-angle approximation ($\theta, \beta, \alpha \ll 1$ radian), $\tan\hat{\alpha} \approx \hat{\alpha}$, and the reduced deflection angle is related to the actual deflection angle by $\hat{\alpha} = (D_{LS}/D_S)\alpha$. Substituting $\xi = D_L\theta$ and $\eta = D_S\beta$ gives the standard lens equation \cite{schneider1992}, expressed as
	\begin{equation}
		\beta = \theta - \frac{D_{LS}}{D_S} \alpha(\theta).
		\label{eq:lens_equation}
	\end{equation}
	
	For a point mass $M$ in the GR weak-field limit, we substitute $\alpha(\xi) = 4GM/(\xi c^2) = 4GM/(D_L\theta c^2)$ from Eq.~(\ref{eq:alpha_gr}) into Eq.~(\ref{eq:lens_equation}). This yields
	\begin{equation}
		\beta = \theta - \frac{4GM}{c^2} \frac{D_{LS}}{D_L D_S} \frac{1}{\theta}.
	\end{equation}
	Multiplying both sides by $\theta$ and rearranging gives the quadratic equation $\theta^2 - \beta\theta - \theta_E^2 = 0$, where we have defined the Einstein angle (or Einstein radius) as
	\begin{equation}
		\theta_E \equiv \sqrt{\frac{4GM}{c^2} \frac{D_{LS}}{D_L D_S}}.
		\label{eq:einstein_radius}
	\end{equation}
	The Einstein angle represents the characteristic angular scale of the gravitational lensing system. It is the radius of the ring formed when the source, lens, and observer are perfectly aligned ($\beta = 0$). The solutions to the lens equation for $\beta \neq 0$ are the two image positions $\theta_\pm = (\beta \pm \sqrt{\beta^2 + 4\theta_E^2})/2$. This derivation confirms that the observable $\theta_E$ depends not only on the lens mass $M$ but also on the distance ratios, highlighting the geometric nature of gravitational lensing as a tool for measuring cosmic distances \cite{schneider1992}.
	
	\subsection{Empirical Validation of the General Relativistic Baseline}
	
	Before proceeding to modified gravity theories, it is essential to verify that the GR framework described above is consistent with observations. The most precise tests of the deflection angle $\alpha$ come from solar system observations, beginning with Eddington's 1919 eclipse expedition \cite{dyson1920}. By measuring the shift of star positions near the Sun's limb, they found a deflection consistent with 1.75 arcseconds, ruling out the Newtonian value of 0.87 arcseconds. Modern very long baseline interferometry (VLBI) has since refined these measurements to an accuracy of approximately 0.01\% by observing quasars as they pass close to the Sun \cite{will2014}.
	
	These empirical results validate the assumption that the refractive index formalism $n(r) = 1 - 2\Phi/c^2$ correctly describes reality in the weak field limit, while simultaneously establishing a rigid baseline that any viable theory of gravity must reproduce the GR deflection angle $\alpha_\text{GR} = 4GM/(bc^2)$ in the high acceleration regime such as the solar system. Consequently, modified gravity theories must be formulated so that deviations become significant only in regimes where GR has not yet been tested with high precision, such as the low acceleration environments of galactic outskirts or cosmological scales. With the geometrical lens equation and the empirical success of GR firmly established, we are now equipped to explore how deviations from the standard Newtonian potential manifest as observable anomalies in the lensing pattern.
	
	\section{Modified Gravity Models and Analytical Results}
	
	In this section, we apply the general deflection formula to several modified gravity models. We first contrast the baseline GR with the distinct constant-deflection behavior of MOND, followed by specific potential modifications in Yukawa and power-law models. For each model, we derive closed-form expressions for the deflection angle where possible, establish the scaling behavior with impact parameter, and crucially, derive the corresponding Einstein radius to link theoretical parameters directly to observables.
	
	\subsection{Modified Newtonian Dynamics (MOND)}
	
	MOND, proposed by Milgrom \cite{milgrom1983}, is fundamentally a non-relativistic modification of dynamics at low accelerations. As such, it cannot make predictions about light propagation without a relativistic extension. The first successful relativistic completion was TeVeS (Tensor-Vector-Scalar theory) developed by Bekenstein \cite{bekenstein2004}, which reproduces MOND dynamics in the non-relativistic limit while providing a framework for gravitational lensing. In TeVeS, the deflection of light differs from our phenomenological prediction due to the interplay between the tensor, vector, and scalar fields. Notably, TeVeS predicts that gravitational lensing in the deep-MOND regime should produce a convergence (surface mass density enhancement) factor of 2 relative to what would be inferred from Newtonian dynamics \cite{zhao2006}. More recent relativistic MOND theories include BIMOND (bi-metric theory) \cite{milgrom2009} and the MOG/STVG theory \cite{moffat2006}. Our phenomenological treatment, which yields $\alpha_\text{MOND} = 2\pi\sqrt{GMa_0}/c^2$, should be understood as an illustrative calculation demonstrating the characteristic constant-deflection behavior, noting that detailed predictions require the full relativistic framework.
	
	In the deep-MOND regime (acceleration $a \ll a_0$, where $a_0 \approx 1.2 \times 10^{-10}$ m/s$^2$), the gravitational acceleration due to a point mass becomes
	\begin{equation}
		g = \frac{\sqrt{GMa_0}}{r},
	\end{equation}
	rather than the standard $g = GM/r^2$. This modification successfully explains galactic rotation curves without invoking dark matter.
	
	The corresponding gravitational potential in the deep-MOND regime is obtained by integrating the acceleration:
	\begin{equation}
		\Phi_{\text{MOND}}(r) = \int g\,dr = \sqrt{GMa_0}\int\frac{dr}{r} = \sqrt{GMa_0}\ln r + C,
	\end{equation}
	where $C$ is an integration constant that does not affect the deflection. Note that this potential grows logarithmically with $r$.
	
	Using the alternative form of the deflection integral, Eq.~(\ref{eq:deflection_potential_direct}), we obtain
	\begin{equation}
		\alpha_{\text{MOND}} = \frac{2}{c^2} \int_{-\infty}^{+\infty} \sqrt{GMa_0} \ln\left(\sqrt{b^2+s^2}\right) \frac{b}{b^2+s^2} ds.
	\end{equation}
	Evaluating this integral (or more simply, using the gradient form with $g = \sqrt{GMa_0}/r$) gives
	\begin{equation}
		\alpha_{\text{MOND}} = \frac{2\pi\sqrt{GMa_0}}{c^2}.
		\label{eq:alpha_mond}
	\end{equation}
	This is a remarkable result where in the deep-MOND regime, the deflection angle is independent of the impact parameter. This contrasts sharply with the GR scaling $\alpha_\text{GR} \propto 1/b$ and provides a distinctive observational signature of MOND. The constancy of the deflection angle arises from the $1/r$ dependence of the gravitational acceleration, which exactly compensates for the geometric $b/r$ factor in the deflection integral.
	
	Because the deflection angle $\alpha_{\text{MOND}}$ is a constant independent of the impact parameter, the lens equation $\beta = \theta - (D_{LS}/D_S)\alpha_{\text{MOND}}$ is linear in $\theta$. For the case of perfect alignment ($\beta = 0$), the Einstein radius is simply
	\begin{equation}
		\theta_E^{\text{MOND}} = \frac{D_{LS}}{D_S} \alpha_{\text{MOND}} = \frac{2\pi\sqrt{GMa_0}}{c^2} \frac{D_{LS}}{D_S}.
		\label{eq:einstein_mond}
	\end{equation}
	This result differs markedly from the GR expression in Eq.~(\ref{eq:einstein_radius}) in its distance dependence where while both scale as $\sqrt{M}$, the MOND Einstein radius lacks the inverse dependence on the lens distance $D_L$ found in the GR formula.
	
	\subsection{Yukawa-Type Modification: Massive Graviton Theories}
	
	A broad class of modified gravity theories, including those predicting a massive graviton or a fifth force, can be modeled by adding a Yukawa-type correction to the Newtonian potential. The modified gravitational potential is written as
	\begin{equation}
		\Phi_Y(r) = -\frac{GM}{r}\left[1 + \alpha_Y e^{-r/\lambda}\right],
		\label{eq:yukawa_potential}
	\end{equation}
	where $\alpha_Y$ is a dimensionless coupling constant determining the strength of the deviation and $\lambda$ is the Compton wavelength of the graviton or the interaction range. If $\alpha_Y > 0$, gravity is enhanced at short distances, whereas if $\alpha_Y < 0$, gravity is suppressed. This form is the standard parametrization for testing deviations from Newtonian gravity in solar system and laboratory experiments \cite{hinterbichler2012,adelberger2003, budhi2025solar}.
	
	To find the deflection angle, we note that the potential is a linear combination of a Newtonian term and a Yukawa term. Thus, the total deflection is the sum of the standard GR deflection and the deflection due to the Yukawa correction. Using Eq.~(\ref{eq:deflection_potential_gradient}) with $\Phi_Y$, we obtain
	\begin{equation}
		\alpha_Y = \frac{4GM}{bc^2} + \alpha_Y \cdot \frac{4GM}{bc^2} e^{-b/\lambda}\left[K_0\left(\frac{b}{\lambda}\right) + \frac{b}{\lambda}K_1\left(\frac{b}{\lambda}\right)\right],
		\label{eq:alpha_yukawa}
	\end{equation}
	where $K_n(z)$ are modified Bessel functions of the second kind. In the limit $b \ll \lambda$, the term in brackets approaches 1, and we recover $\alpha_Y \to (4GM/bc^2)(1 + \alpha_Y)$, consistent with an effective gravitational constant $G_{\text{eff}} = G(1+\alpha_Y)$.
	
	To determine the observable Einstein radius $\theta_E^Y$, we must solve the lens equation with this deflection angle. Substituting $\alpha_Y$ from Eq.~(\ref{eq:alpha_yukawa}) into Eq.~(\ref{eq:lens_equation}) with $\beta = 0$ gives
	\begin{equation}
		\theta = \frac{D_{LS}}{D_S} \cdot \frac{4GM}{c^2 D_L \theta} \left\{1 + \alpha_Y e^{-D_L\theta/\lambda}\left[K_0\left(\frac{D_L\theta}{\lambda}\right) + \frac{D_L\theta}{\lambda}K_1\left(\frac{D_L\theta}{\lambda}\right)\right]\right\}.
	\end{equation}
	This equation is transcendental and must be solved numerically. The behavior is determined by the coupling $\alpha_Y$ where a positive $\alpha_Y$ predicts a larger Einstein ring than GR, while a negative $\alpha_Y$ predicts a smaller one.
	
	\subsection{Power-Law Modification: \texorpdfstring{$f(R)$}{f(R)} Gravity and Scalar-Tensor Theories}
	
	Power-law modifications to the gravitational potential arise naturally in $f(R)$ gravity theories, where the Einstein-Hilbert action is extended to include non-linear functions of the Ricci scalar $R$. In these theories, an additional scalar degree of freedom (the scalaron) mediates a fifth force that modifies the gravitational potential \cite{sotiriou2010,defelice2010,budhi2025inflation}. A crucial feature of $f(R)$ gravity is the chameleon mechanism \cite{khoury2004, budhi2025ps, budhi2025sing}, which renders the additional force environment-dependent. For gravitational lensing, this means that the effective potential experienced by light may differ from that governing non-relativistic particle motion. The phenomenological potential we consider approximates the behavior in the unscreened regime. Complete lensing derivations within $f(R)$ cosmology are presented by Schmidt \cite{schmidt2008} and Terukina et al.\ \cite{terukina2011}, who found that lensing constraints tightly limit deviations from general relativity.
	
	We consider the phenomenological potential
	\begin{equation}
		\Phi_{PL}(r) = -\frac{GM}{r}\left[1 + \epsilon\left(\frac{r_0}{r}\right)^n\right],
		\label{eq:powerlaw_potential}
	\end{equation}
	where $\epsilon$ is a dimensionless coupling parameter, $r_0$ is a characteristic length scale, and $n > 0$ determines the power of the correction. For $n = 1$, this form arises in certain $f(R)$ gravity models, whereas for $n = 2$, it corresponds to the next-to-leading order correction in some effective field theory treatments.
	
	Using Eq.~(\ref{eq:deflection_potential_direct}) with $\Phi_{PL}$, the total deflection angle is
	\begin{equation}
		\alpha_{PL} = \frac{4GM}{bc^2}\left[1 + \epsilon(n+1)\left(\frac{r_0}{b}\right)^n \frac{\sqrt{\pi}\,\Gamma\left(\frac{n+2}{2}\right)}{2\Gamma\left(\frac{n+3}{2}\right)}\right].
		\label{eq:alpha_powerlaw}
	\end{equation}
	The coefficient involving gamma functions simplifies for specific values of $n$. For $n = 1$, we have $\Gamma(3/2) = \sqrt{\pi}/2$ and $\Gamma(2) = 1$, giving a correction factor of $\pi\epsilon r_0/(2b)$. For $n = 2$, we have $\Gamma(2) = 1$ and $\Gamma(5/2) = 3\sqrt{\pi}/4$, giving a correction factor of $2\epsilon r_0^2/b^2$.
	
	Inserting the power-law deflection angle into the lens equation ($\beta = 0$) yields a polynomial equation for $\theta$:
	\begin{equation}
		\theta^{n+2} - \theta_{E,\text{GR}}^2 \theta^n - \epsilon (n+1) C_n \theta_{E,\text{GR}}^2 \left(\frac{r_0}{D_L}\right)^n = 0,
		\label{eq:theta_powerlaw}
	\end{equation}
	where $C_n = \sqrt{\pi}\,\Gamma\left(\frac{n+2}{2}\right) / \left[2\Gamma\left(\frac{n+3}{2}\right)\right]$ and $\theta_{E,\text{GR}}$ is given by Eq.~(\ref{eq:einstein_radius}). For $\epsilon = 0$, the solution reduces to the GR Einstein radius. For non-zero $\epsilon$, the positive root of this equation defines the modified Einstein radius $\theta_E^{PL}$. Since the correction term scales as $\theta^{-n}$, it becomes dominant at small $\theta$ (or small $b$). Consequently, for a given lens mass, a power-law modification with $\epsilon > 0$ predicts an Einstein radius that is larger than the GR prediction.
	
	\subsection{Summary of Scaling Relations}
	
	For clarity of presentation, we summarize the scaling relations between deflection angle $\alpha(b)$ and impact parameter $b$, and the resulting Einstein radius $\theta_E$ for the various models considered in Table~\ref{tab:scaling}. These distinct scaling relations provide, in principle, a means of distinguishing between different gravitational theories through observations of lensing by systems with varying impact parameters.
	
	\begin{table}[htbp]
		\caption{Scaling relations between deflection angle $\alpha(b)$ and impact parameter $b$, and the Einstein radius $\theta_E$ for different gravity models.}
		\label{tab:scaling}
		\begin{ruledtabular}
			\begin{tabular}{lccc}
				\hline
				Model & Potential $\Phi(r)$ & Scaling $\alpha(b)$ & Einstein Radius $\theta_E$ \\
				\hline
				GR Weak-Field & $-GM/r$ & $\propto 1/b$ & $\theta_E \propto \sqrt{M/D_L}$ \\
				MOND (deep) & $\sqrt{GMa_0}\ln r$ & constant & $\theta_E \propto \sqrt{M}$ \\
				Yukawa & $-(GM/r)[1 + \alpha_Y e^{-r/\lambda}]$ & $\propto (1+\alpha_Y \cdots)/b$ & Transcendental \\
				Power-law & $-(GM/r)[1 + \epsilon(r_0/r)^n]$ & $\propto 1/b + \epsilon/b^{n+1}$ & Polynomial root (Eq.~\ref{eq:theta_powerlaw}) \\
				\hline
			\end{tabular}
		\end{ruledtabular}
	\end{table}
	
	\section{Computational Implementation}
	
	The computational approach employed in this work is designed to be accessible to undergraduates with basic programming experience. The ray equation $d^2\mathbf{r}/ds^2 = \nabla n$ is solved numerically using standard scientific computing libraries available in Python, a widely-used programming language in scientific research. Specifically, we utilize the \texttt{scipy.integrate.odeint} function, which implements robust numerical integration algorithms suitable for systems of ordinary differential equations. This choice is advantageous for teaching because the function handles the technical details of variable step-size integration internally, allowing students to focus on the physics formulation rather than the intricacies of numerical algorithms. Furthermore, the Python ecosystem provides excellent visualization capabilities through the \texttt{matplotlib} library, enabling students to generate publication-quality figures with minimal effort.
	
	The implementation proceeds by formulating the second-order ray equation as a system of first-order differential equations. We introduce the velocity variable $\mathbf{v} = d\mathbf{r}/ds$ and write
	\begin{align}
		\frac{d\mathbf{r}}{ds} &= \mathbf{v}, \\
		\frac{d\mathbf{v}}{ds} &= \nabla n(\mathbf{r}).
	\end{align}
	For a spherically symmetric refractive index $n(r)$, the gradient is $\nabla n = (dn/dr)\hat{\mathbf{r}}$, where $\hat{\mathbf{r}} = \mathbf{r}/r$. In Cartesian coordinates, the equations become
	\begin{align}
		\frac{dx}{ds} &= v_x, \quad \frac{dy}{ds} = v_y, \quad \frac{dz}{ds} = v_z, \\
		\frac{dv_x}{ds} &= \frac{dn}{dr}\frac{x}{r}, \quad \frac{dv_y}{ds} = \frac{dn}{dr}\frac{y}{r}, \quad \frac{dv_z}{ds} = \frac{dn}{dr}\frac{z}{r}.
	\end{align}
	By appropriate choice of coordinate system, the problem can be reduced to two dimensions without loss of generality. Taking the lens to lie at the origin and the incoming ray to travel initially in the $z$-direction with impact parameter $b$ in the $x$-direction, the trajectory remains in the $x$-$z$ plane, and we need only integrate the four equations for $x$, $z$, $v_x$, and $v_z$.
	
	Initial conditions are set at a large distance $z_0$ upstream of the lens, where the gravitational potential is negligible:
	\begin{align}
		x(0) &= b, \quad z(0) = -z_0, \\
		v_x(0) &= 0, \quad v_z(0) = 1.
	\end{align}
	The integration proceeds until the ray reaches a distance $z_f$ downstream of the lens, where the deflection angle is extracted from the final direction vector via the relation
	\begin{equation}
		\alpha = \arctan\left(\frac{v_x(s_f)}{v_z(s_f)}\right) \approx \frac{v_x(s_f)}{v_z(s_f)},
	\end{equation}
	where the approximation holds for small deflection angles.
	
	To ensure numerical stability, the solver allows for the specification of relative and absolute error tolerances (\texttt{rtol} and \texttt{atol}). For the potentials considered in this work, default tolerances ($\texttt{rtol} = 1.49 \times 10^{-8}$, $\texttt{atol} = 1.49 \times 10^{-8}$) provide sufficient precision for instructional purposes. Students are encouraged to verify convergence by comparing results obtained with different tolerance settings or by comparing numerical results with analytical predictions where available.
	
	It is important to note the distinction between the physical deflection angle $\alpha$ and the observed image position $\theta$. While the analytical derivations in Section III rigorously account for the angular diameter distances ($D_L, D_S, D_{LS}$) via the lens equation, our numerical simulations are performed in a simplified normalized unit system. We adopt units where $G = M = c = 1$ and, for the purpose of comparing the intrinsic scaling of deflection, we set the geometric factor $D_{LS}/(D_L D_S)$ to unity. To regain physical intuition, we note that in these normalized units, a deflection angle of order unity corresponds to the strong-field regime near compact objects, whereas actual astrophysical lensing involves deflections of milliarcseconds, corresponding to values of order $10^{-8}$--$10^{-6}$ in these units. For reference, the GR deflection of light grazing the Sun, one of the largest weak-field deflections known, is 1.75 arcseconds ($\approx 8.5 \times 10^{-6}$ rad). Furthermore, this normalization makes the deflection effects visually conspicuous in the ray-tracing plots, allowing students to clearly observe the distinct bending patterns predicted by each gravity model. In physical units, the deflections are too small to be discerned in a single plot, but the normalized units amplify the geometric features without altering the underlying physics, thereby enhancing the instructional value of the simulations.
	
	Finally, we validate the implementation by ensuring that the code reproduces the analytical GR deflection angle and captures the correct scaling behavior for the modified gravity models discussed in the previous sections. The complete Python code used to generate all figures in this paper is available in the public repository \cite{github_repo}, allowing students and instructors to run, modify, and extend the simulations directly.
	
	\section{Results and Discussion}
	
	\begin{figure}[htbp]
		\centering
		\includegraphics[width=0.8\linewidth]{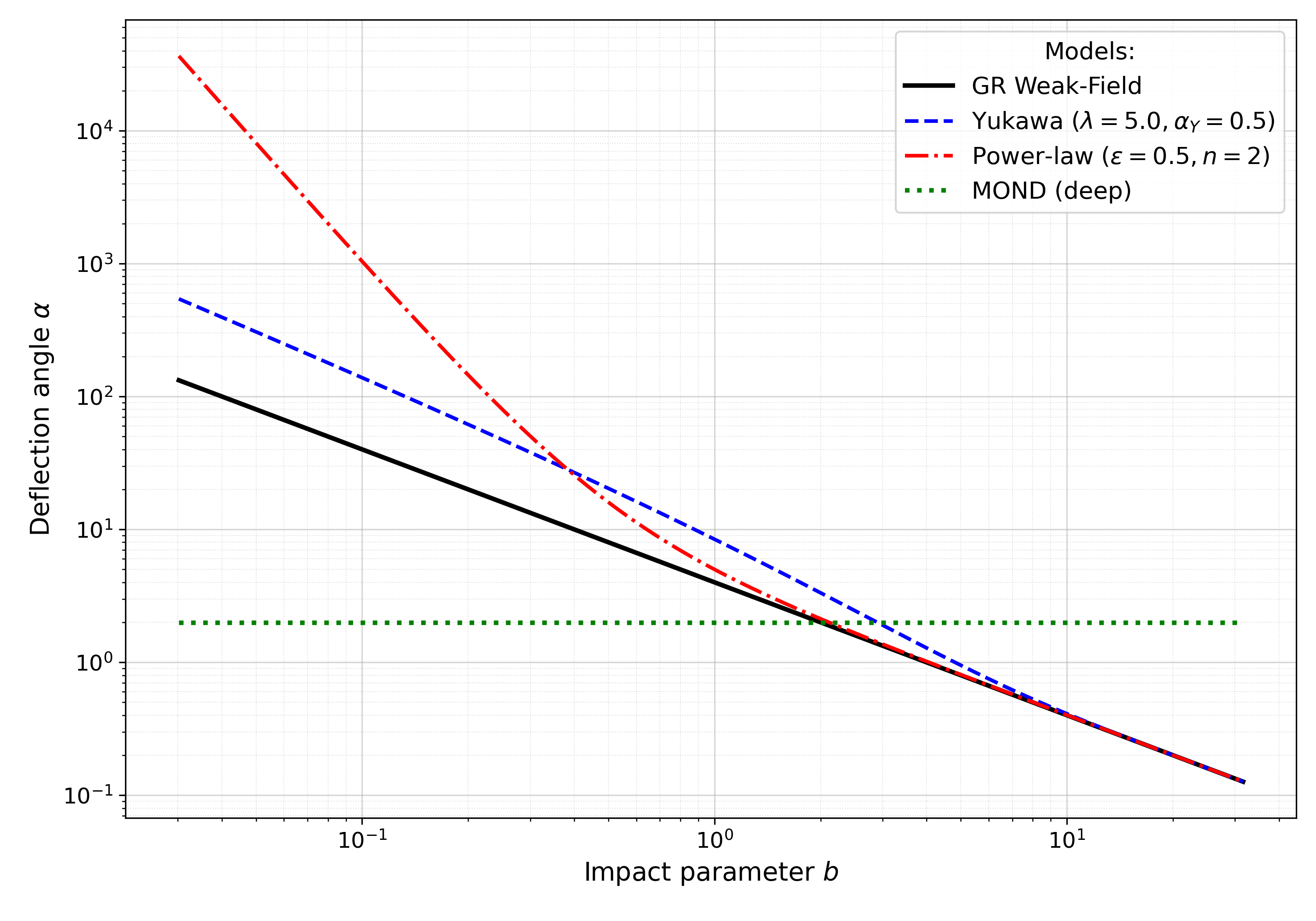}
		\caption{Deflection angle as a function of impact parameter for various gravity models. The GR weak field case exhibits the characteristic $\alpha \propto 1/b$ scaling (solid black line). The Yukawa modification (dashed blue) with positive coupling $\alpha_Y$ shows enhanced deflection. The power law model (dot-dashed red) displays even stronger enhancement at small $b$. The MOND model (dotted green) predicts a constant deflection angle independent of $b$, contrasting sharply with the GR scaling. Parameters used are $\lambda = 5$, $\alpha_Y = 0.5$, $n = 2$, $\epsilon = 0.5$, and $a_0 = 0.1$ in the normalized units described in Section IV.}
		\label{fig:alpha_vs_b}
	\end{figure}
	
	\begin{figure}[htbp]
		\centering
		\includegraphics[width=0.95\linewidth]{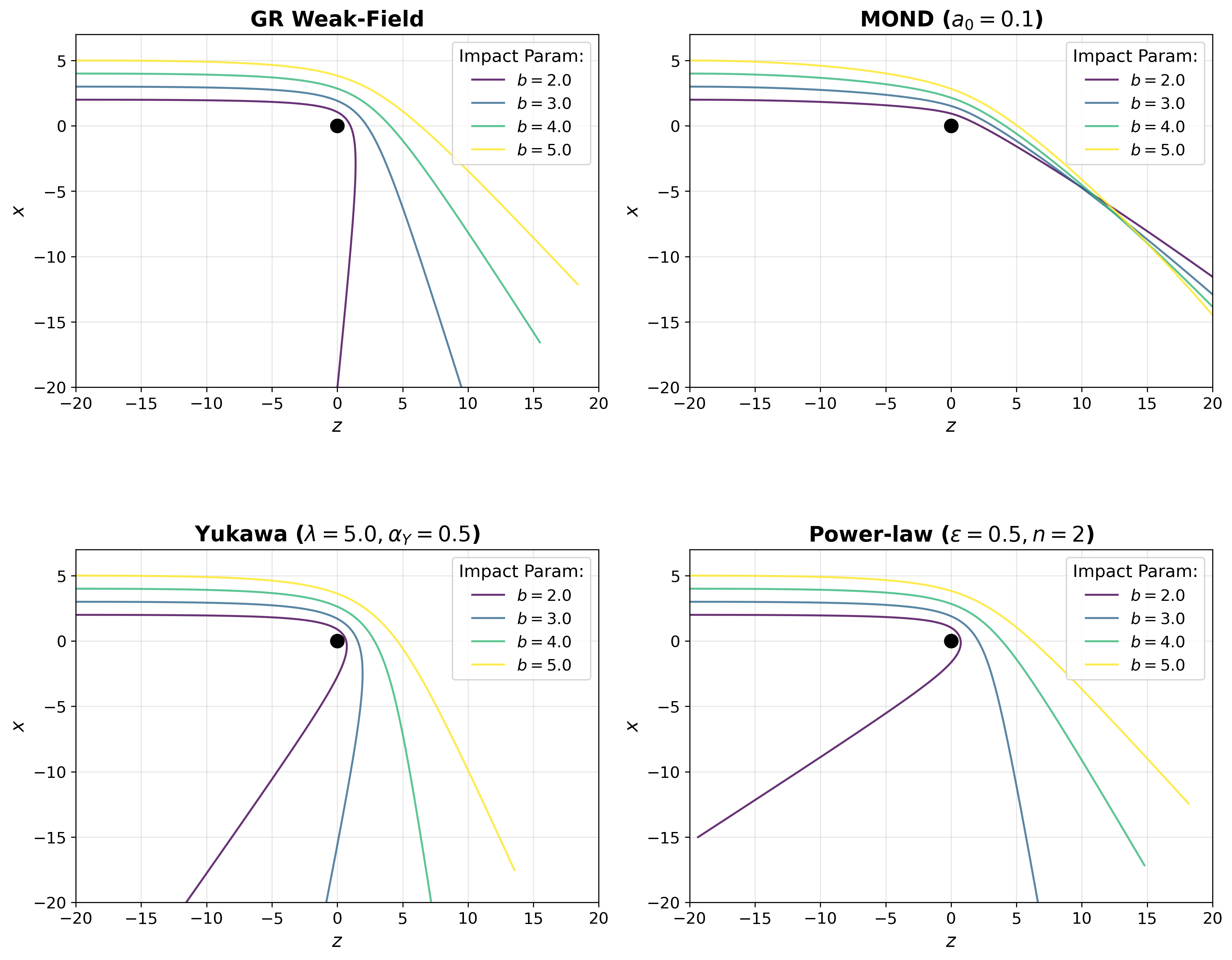}
		\caption{Ray tracing simulations showing gravitational lensing trajectories for different gravity models. Each panel displays bundles of rays with impact parameters $b = 2.0, 3.0, 4.0, 5.0$ (in normalized units) passing through the gravitational field. Top left shows GR weak field gravity with characteristic $1/b$ deflection scaling. Top right shows MOND (deep regime) where all rays are deflected by the same asymptotic angle regardless of $b$. Although the trajectories for different $b$ appear to have different curvatures and final transverse positions, their final direction vectors are identical, reflecting the constant deflection angle predicted by Eq.~(\ref{eq:alpha_mond}). Bottom left shows the Yukawa modification with enhanced bending. Bottom right shows the power law modification exhibiting the strongest deflection near the lens. The black dot indicates the lens position at the origin.}
		\label{fig:ray_tracing}
	\end{figure}
	
	\begin{figure}[htbp]
		\centering
		\includegraphics[width=0.95\linewidth]{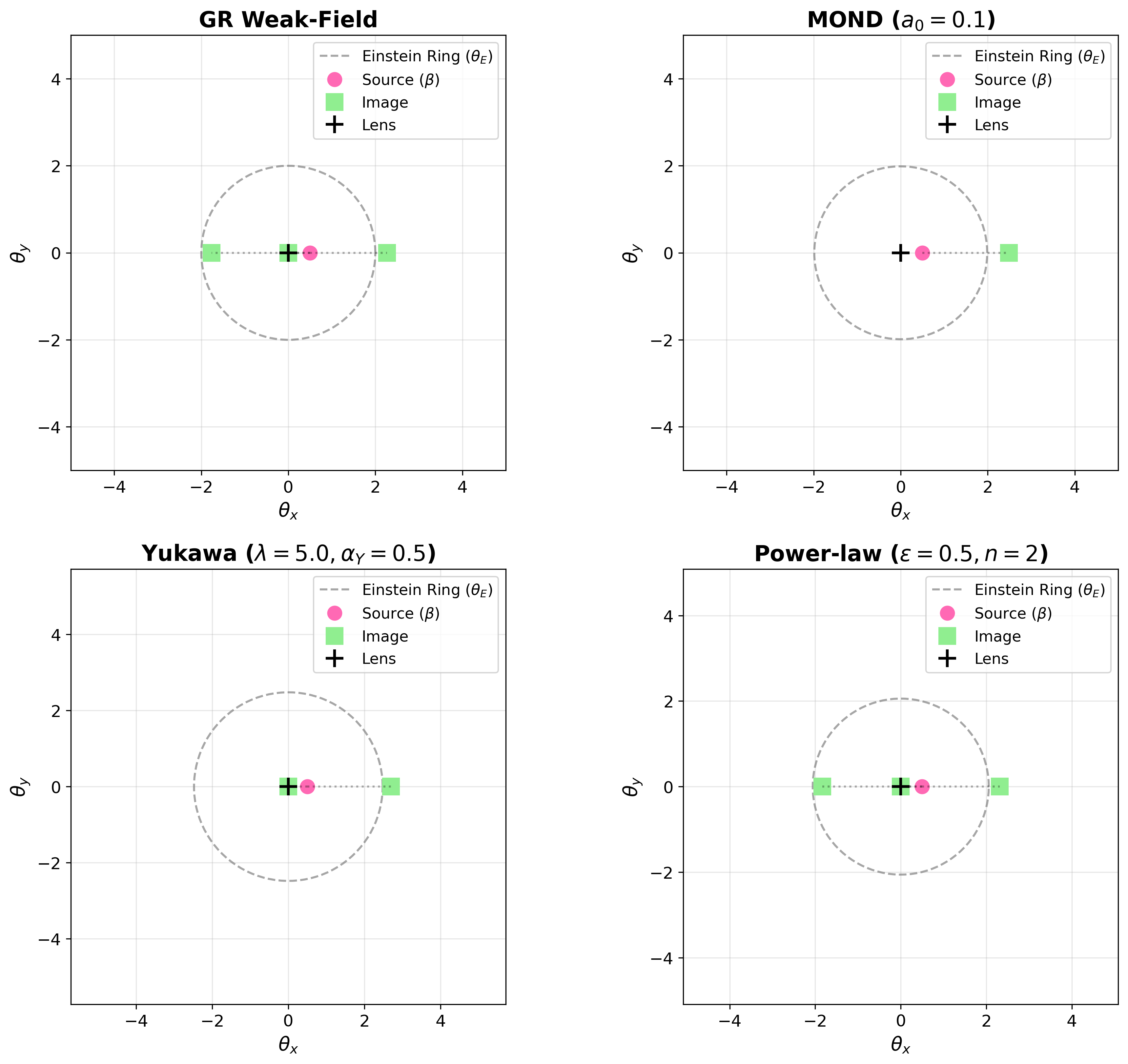}
		\caption{Einstein ring geometry and image formation for each model. The panels show the Einstein ring (dashed circle) and the image positions (squares) for a source located at $\beta = 0.5$  (in normalized units, corresponding to radians under the convention $G=M=c=1$). Differences in ring size and image separation highlight the distinct lensing signatures. All quantities are in normalized units.}
		\label{fig:einstein_ring}
	\end{figure}
	
	Figure~\ref{fig:alpha_vs_b} shows the scaling of the deflection angle $\alpha$ with impact parameter $b$ for all models. The GR baseline exhibits the characteristic $1/b$ dependence, where the deflection increases as the light ray passes closer to the lens. The Yukawa type correction with positive coupling $\alpha_Y > 0$ enhances gravity at small scales, producing a larger deflection that transitions to the GR value at large distances. The power law modification, typical of $f(R)$ gravity, enhances the deflection significantly at small impact parameters, suggesting that strong lensing events near galactic centers could probe higher order curvature terms. The most striking behavior appears in the deep MOND regime, which predicts a deflection angle independent of the impact parameter. This constancy arises from the $1/r$ scaling of the gravitational acceleration in this potential, which exactly compensates for the geometric $b/r$ factor in the deflection integral.
	
	These theoretical scaling relations translate directly into observable geometric features, as demonstrated by the ray tracing simulations in Fig.~\ref{fig:ray_tracing}. The GR trajectories show the familiar pattern of asymptotic bending where curvature decreases with distance. The Yukawa model produces trajectories similar to GR at large $b$ but curves more sharply near the lens due to the $\alpha_Y$ term. The power law model shows more pronounced bending close to the lens, consistent with the divergent nature of the correction term. The most profound geometric distinction appears in the MOND case, where all rays are deflected by the same asymptotic angle regardless of their distance from the lens centre, leading to image formation that differs qualitatively from GR. In the ray-tracing plot, the trajectories for different impact parameters exhibit different curvatures and final transverse positions, which might be mistaken for varying deflection angles. However, the final direction vectors are identical, as confirmed numerically by computing $\arctan(v_x/v_z)$ at the end of the integration. This constancy of the asymptotic deflection angle is a direct consequence of the $1/r$ scaling of the gravitational acceleration in the deep-MOND regime. Figure~\ref{fig:einstein_ring} further elucidates this by comparing the Einstein ring geometry for each model. The Yukawa and power law models predict larger Einstein rings than GR for the same mass parameter due to the positive coupling constants used in our simulation.
	
	Despite the theoretical variety offered by these modified models, observational data place stringent constraints on their viability. The most precise tests originate from the Solar System, where VLBI measurements of quasars near the Sun have confirmed the GR prediction to within 0.01\% \cite{will2014}. This accuracy requires that any viable modification, such as the Yukawa coupling $\alpha_Y$ or power law parameter $\epsilon$, must be heavily suppressed in high acceleration environments. On galactic and cosmological scales, strong lensing surveys like SLACS \cite{bolton2006} find excellent agreement between measured Einstein radii and GR predictions assuming standard stellar and dark matter distributions. The Bullet Cluster observation \cite{clowe2006} poses a significant challenge to theories like MOND that attempt to replace dark matter with modified gravity unless additional non baryonic fields are introduced. These observations reinforce the point that while exploring alternative theories is intellectually stimulating, GR combined with the dark matter paradigm remains the most robust description of gravitational lensing phenomena.
	
	The observational constraints can be translated into quantitative bounds on the parameters of the modified gravity models considered in this work. In the deep-MOND regime, the deflection angle $\alpha_{\text{MOND}} = 2\pi\sqrt{GMa_0}/c^2$ is independent of $b$. For a point mass $M$, the ratio $\alpha_{\text{MOND}}/\alpha_{\text{GR}} = (\pi/2) \sqrt{a_0 b^2/(GM)}$. At solar system scales ($b \sim R_\odot$, $M = M_\odot$), this ratio is of order $10^{-5}$, consistent with the null result of VLBI. However, for galactic scales ($b \sim 10$ kpc, $M \sim 10^{11} M_\odot$), the ratio becomes of order unity, allowing MOND to potentially explain galaxy rotation curves without dark matter. The Bullet Cluster observation \cite{clowe2006} poses a more severe challenge in that the separation between the X-ray gas and the gravitational lensing signal requires a significant mass component that does not interact collisionally, which in MOND frameworks necessitates additional fields (e.g., neutrinos) and remains a subject of active debate.
	
	For the Yukawa-type potential, solar system VLBI measurements at an impact parameter $b \sim R_\odot$ (the solar radius) require that the fractional deviation from GR satisfies $|\alpha_Y| e^{-R_\odot/\lambda} [K_0(R_\odot/\lambda) + (R_\odot/\lambda)K_1(R_\odot/\lambda)] \lesssim 10^{-4}$, given the 0.01\% accuracy of deflection measurements \cite{will2014}. For $\lambda \gg R_\odot$, the modified Bessel functions behave as $K_0(z) \approx -\ln(z/2) - \gamma$ and $K_1(z) \approx 1/z$, leading to the approximate bound $|\alpha_Y| \lesssim 10^{-4} / (R_\odot/\lambda)^2$ for $\lambda \gg R_\odot$. For $\lambda \sim R_\odot$, the bound becomes $|\alpha_Y| \lesssim 10^{-4}$. Thus, any significant deviation ($|\alpha_Y| \sim 1$) would require the interaction range $\lambda$ to be either much larger than astronomical scales (where the exponential suppression becomes weak) or much smaller than solar system scales (where the Yukawa correction is exponentially suppressed). This dichotomy illustrates why fifth force searches often focus on laboratory or cosmological scales.
	
	For the power-law $f(R)$ type modification, the fractional deviation from GR scales as $\epsilon (n+1) C_n (r_0/b)^n$. VLBI observations at $b \sim R_\odot$ constrain $\epsilon (r_0/R_\odot)^n \lesssim 10^{-4} / [(n+1)C_n]$. For $n=1$ and taking $r_0$ to be of order the solar radius, this gives $\epsilon \lesssim 10^{-4}$. For $n=2$, the constraint is $\epsilon (r_0/R_\odot)^2 \lesssim 10^{-4}$. Therefore, if the characteristic scale $r_0$ is of galactic size ($\sim 10$ kpc), the factor $(r_0/R_\odot)^2 \sim 10^{22}$ forces $\epsilon$ to be astronomically small ($\epsilon \lesssim 10^{-26}$), indicating that any detectable effect in galaxy-scale lensing would require a much larger $r_0$ or a different $n$.

	These quantitative estimates, summarized in Table~\ref{tab:parameter_bounds}, demonstrate how current observations constrain the parameter space of modified gravity models. While a full statistical analysis is beyond the scope of this instructional treatment, the exercise illustrates the essential connection between theory and experiment, and provides students with a concrete example of how observational data can be used to test alternative theories.
	
	\begin{table}[htbp]
		\caption{Approximate observational bounds on modified gravity parameters from solar system VLBI measurements \cite{will2014} and galactic lensing surveys \cite{bolton2006}. The bounds assume the characteristic scale $r_0$ is of order the solar radius $R_\odot$ for the power-law model, and $\lambda \sim R_\odot$ for the Yukawa model. For other scales, the bounds scale accordingly.}
		\label{tab:parameter_bounds}
		\begin{ruledtabular}
			\begin{tabular}{lcc}
				\hline
				Model & Parameter & Approximate Bound \\
				\hline
				MOND & $a_0$ & $1.2 \times 10^{-10}$ m/s$^2$ (fitted) \\
				Yukawa ($\lambda \sim R_\odot$) & $|\alpha_Y|$ & $\lesssim 10^{-4}$ \\
				Yukawa ($\lambda \gg R_\odot$) & $|\alpha_Y| (R_\odot/\lambda)^2$ & $\lesssim 10^{-4}$ \\
				Power-law ($n=1$, $r_0 \sim R_\odot$) & $|\epsilon|$ & $\lesssim 10^{-4}$ \\
				Power-law ($n=2$, $r_0 \sim R_\odot$) & $|\epsilon|$ & $\lesssim 10^{-4}$ \\
				Power-law ($n=1$, $r_0 \sim 10$ kpc) & $|\epsilon|$ & $\lesssim 10^{-26}$ \\
				\hline
			\end{tabular}
		\end{ruledtabular}
	\end{table}
	
	The optical mechanical analogy thus provides a rigorous yet accessible framework for teaching gravitational lensing and exploring modified gravity theories. By integrating analytical derivations, numerical simulations, and observational context, this approach offers a holistic educational experience that prepares students for the complexities of modern gravitational physics.
	
	\section{Conclusions}
	
	We have presented a comprehensive instructional framework that transforms gravitational lensing from a specialized topic in general relativity into an accessible application of undergraduate mechanics and optics. By rigorously establishing the optical mechanical analogy and the effective refractive index $n(r) = 1 - 2\Phi(r)/c^2$, we have clarified the fundamental distinction between Newtonian gravity and general relativity, resolving a common conceptual confusion for students. This framework reproduces the standard GR predictions, including the $1/b$ scaling of the deflection angle and the Einstein radius, while also serving as a versatile platform for exploring modified gravity theories. Through the analysis of MOND, Yukawa type corrections, and power law $f(R)$ potentials, we have demonstrated how distinct theoretical models manifest as unique observational signatures. The most striking example is the deep MOND regime, which predicts a constant deflection angle independent of the impact parameter, contrasting sharply with the GR scaling.
	
	The analytical predictions have been validated through numerical ray tracing simulations, which confirm the distinct deflection patterns for each model. The GR trajectories exhibit asymptotic bending that decreases with distance, whereas the Yukawa model shows enhanced curvature near the lens, the power law modification produces even stronger bending at small impact parameters, and the MOND case displays the unique feature of equal deflection for all rays regardless of their distance from the lens center. These distinct signatures translate directly into observable differences in Einstein ring sizes and image formations, as summarized in the scaling relations of Table~\ref{tab:scaling}. Students can verify these results computationally, thereby bridging the gap between abstract mathematical formalism and tangible physical phenomena.
	
	Observational data from solar system VLBI measurements \cite{will2014}, strong lensing surveys such as SLACS \cite{bolton2006}, and the Bullet Cluster \cite{clowe2006} place stringent constraints on modified gravity theories. As shown in Table~\ref{tab:parameter_bounds}, these constraints force the deviation parameters $\alpha_Y$ and $\epsilon$ to be smaller than about $10^{-4}$ at solar system scales, while allowing order-unity deviations on galactic scales for MOND. Students can use the derived scaling relations to estimate how the bounds scale with the interaction range $\lambda$ or the characteristic length $r_0$, thereby gaining insight into the interplay between theory and experiment. These empirical realities reinforce the educational value of our framework. By stripping away the requirement for advanced differential geometry, we enable advanced undergraduates to engage meaningfully with frontier questions regarding dark matter, dark energy, and modified gravity. This framework does not replace the need for rigorous relativistic study, but it provides a crucial conceptual and computational stepping stone that empowers students to approach the full machinery of general relativity with intuition and confidence.
	

\end{document}